\documentclass[final,5p,times,twocolumn,numeric]{elsarticle}

\usepackage{amsmath}
\usepackage{graphicx}
\usepackage{amsfonts}
\usepackage{latexsym}
\usepackage{bbold}
\usepackage{wasysym}
\usepackage{calligra}
\usepackage{float}
\usepackage{ulem}
\usepackage{inputenc}
\usepackage{xspace}
\usepackage{url}
\usepackage{epstopdf}
\usepackage{tikz}
\usepackage{amssymb}
\usepackage{enumitem}

\usepackage{appendix}

\usepackage{amsmath}
\usepackage{amssymb}
\usepackage{graphicx}
\usepackage{amsfonts}
\usepackage{latexsym}
\usepackage{bbold}
\usepackage{wasysym}
\usepackage{calligra}
\usepackage{float}
\usepackage{ulem}
\usepackage{inputenc}
\usepackage{xspace}
\usepackage{url}
\usepackage{epstopdf}
\usepackage{tikz}
\usepackage{amsthm}
\usepackage{soul}

\newcommand{\mc}{\mathcal{M}}
\newcommand{\y}{\upsilon}
\newcommand{\be}{\begin{equation}}
\newcommand{\ee}{\end{equation}}
\newcommand{\beq} {\begin{equation}}
\newcommand{\eeq} {\end{equation}}
\newcommand{\ba}{\begin{eqnarray}}
\newcommand{\ea}{\end{eqnarray}}

\usepackage{amssymb}
\usepackage{lipsum}

\journal{Physics Letters B}

\begin{document}

\begin{frontmatter}



\title{Novel Regge-like trajectories for spinning, dilating, hadronic particles}


\author[first,f]{Damianos Iosifidis}
\affiliation[first]{organization={Scuola Superiore Meridionale, Largo San Marcellino 10, 80138 Napoli, Italy},
}

\affiliation[f]{organization={INFN– Sezione di Napoli, Via Cintia, 80126 Napoli, Italy}
}

\begin{abstract}
We study the of motion of a spinning, dilating particle with hadronic properties moving on a generic geometric background including curvature, torsion, and nonmetricity. In particular, we discuss generalized spin supplementary conditions and also introduce the concept of a shear supplementary condition. Using these, we investigate the evolution of the dynamical mass of the microstructured test  body and the cases where the latter is a constant of motion. In general, we find novel Regge-like trajectories relating the mass to the dilation and/or the shear currents of hypermomentum. This means that for particles with hadronic properties, the rest mass is not a constant of motion in general.
\end{abstract}




\end{frontmatter}




\section{Introduction}
\label{introduction}

The standard description of hadronic interactions via Quantum Chromodynamics (QCD) relies on the gauge symmetry of color-charged gluons, characterized by the  $SU(3)$ group.Complementing this microscopic view, the observed macroscopic properties of hadrons—specifically their arrangement into Regge trajectories where the squared mass 
 scales linearly with angular momentum ($\mathcal{M}^{2}\propto J$)
—suggest an underlying structural symmetry that transcends the point-particle approximation. Early foundational work by Ne’eman and  Šijački \cite{NeemanSijacki1979} proposed that these "tower" states of hadrons are best described by the unitary irreducible representations (unirrep) of the universal covering of the special linear group $\overline{SL}(4,R)$. By incorporating dilations, this symmetry is extended to the general linear group $GL(4,R)$, which serves as the gauge group of Metric-Affine Gravity (MAG)\cite{hehl1995metric}, developed by pioneering works of Hehl and collaborators \cite{hehl1995metric,Hehl:1976my,hehl1976hypermomentum,hehl1981metric,hehl1999metric}.

In this framework, the transition from the Riemannian structure of General Relativity to the non-Riemannian structure of MAG  provides, along with curvature, two new geometric entities; torsion, and nonmetricity. For minimally coupled matter it is generally accepted that  while torsion is coupled to the intrinsic spin of particles, nonmetricity is coupled to the  the symmetric parts of hypermomentum, namely dilation and shear. The inclusion of the shear generator is particularly significant, as it corresponds to the volume-preserving deformations of the hadronic "bag" or string \cite{NeemanSijacki1979}.

Furthermore, Šijački \cite{SIJACKI1982297} showed that the confinement of quarks can be naturally explained by passing from the Riemannian  to the (non-Riemannian) Metric-Affine  Geometry. Within this dual geometric-gauge framework, the gluon field strength is reinterpreted as a dynamical component of the affine connection, effectively identifying the nonmetricity of the non-Riemannian spacetime manifold at the hadronic scale as the geometric manifestation of color-confining dynamics. This also suggests that the non-metric structure of spacetime, manifests itself in extremely high energies, e.g. in situations like quark-gluon plasma.

More generally, this particle microstructure-extended geometry  association clearly suggests that in order to probe the full non-Riemannian structure of spacetime, matter with microstructure must be used. Reversely, the micro-properties of matter are revealed when the matter is placed in generalized geometric backgrounds admitting torsion and nonmetricity. Indeed, this conclusion is supported by several investigations \cite{Puetzfeld:2007hr,Puetzfeld:2007ye}. On a kinematical level one may wonder what are the consequences of this matter-geometry interaction. In particular, compared to the Riemannian picture, how is the motion of a mictrostructured particle affected by the enriched external geometry? In this note we study some of the implications of such couplings. In particular, starting from the equations of motion of microstructured test particles in non-Riemannian backgrounds, we investigate under what circumstances the mass is a constant of motion. In addition we obtain novel Regge-like trajectories that relate the mass to the dilation and shear charges of the particle respectively.

\section{Geometry and Motion}

We shall consider a generic n-dimensional geometric background endowed with a metric $g_{\mu\nu}$ and a general linear connection $\Gamma^{\lambda}{}_{\mu\nu}$. In this metric-affine space we  define the torsion, curvature and nonmetricity according to
\begin{subequations}
\begin{align}
S_{\mu\nu}^{\;\;\;\lambda}&:=\Gamma^{\lambda}_{\;\;\;[\mu\nu]}
\;\;, \;\; \\
R^{\mu}_{\;\;\;\nu\alpha\beta}&:= 2\partial_{[\alpha}\Gamma^{\mu}_{\;\;\;|\nu|\beta]}+2\Gamma^{\mu}_{\;\;\;\rho[\alpha}\Gamma^{\rho}_{\;\;\;|\nu|\beta]} \;\;,\label{R} 
  \\
Q_{\alpha\mu\nu}&:=- \nabla_{\alpha}g_{\mu\nu}
\end{align}
\end{subequations}
respectively.
This is the geometric arena in which motion of particles takes place. 

Let us introduce the kinematical characteristics of the particle (or body). If $x^{\alpha}=x^{\alpha}(\lambda)$ is the representative worldline of the particle, $\lambda$ being an arbitrary parametrization of the curve, then the un-normalized and normalized velocity fields are respectively defined through
\beq
\y^{\alpha}:=\frac{d x^{\alpha}}{d \lambda}\;\;, \;\; u^{\alpha}:=\frac{d x^{\alpha}}{d \tau}
\eeq
 where $\tau$ denotes proper time parametrization and $u^{\mu}u_{\mu}=-1$ by construction. The directional derivative along $\y^{\alpha}$ is defined by
\beq
\frac{D}{d \lambda}:=\y^{\alpha}\nabla_{\alpha}
\eeq
and may act on tensors of arbitrary rank. For simplicity we will sometimes just use an over-dot  to indicate this derivative, for example 
\beq
\dot{H}^{\mu\nu}:=\frac{D H^{\mu\nu}}{d \lambda}=\y^{\alpha}\nabla_{\alpha}H^{\mu\nu}
\eeq

The equations of motion of spinning particles under the influence of Riemannian  gravitational fields were derived in pioneering works of Mathisson, Papapetrou and Dixon \cite{Mathisson:1937zz,Papapetrou:1951pa,Dixon:1974xoz}. Every-since, countless papers have been written on the matter. For a nice chronological review of the equations of motion in General Relativity, see \cite{Puetzfeld:2007hr}. In these works only the spin part of the particle's microstructure is taken into account. The next step is to consider also the hadronic properties of matter and study the motion of microstructured bodies in generalized geometric backgrounds. Such investigations have been initiated in \cite{Puetzfeld:2007hr,Puetzfeld:2007ye}.

In \cite{Iosifidis:2023eom},  by integrating the generalized MAG conservation laws, the equations of motion of a spinning particle with hadronic properties (i.e. including also the dilation and shear charges) were derived over the most general geometric background. The generalization using  a Lagrangian description was given in \cite{Iosifidis:2025ano}. In full generality, the evolution equations of the micro-structured body, read
\beq
\frac{D P_{\mu}}{d\lambda}=-2 S_{\mu\alpha\beta}\upsilon^{\alpha}P^{\beta}+\frac{1}{2}H^{\alpha\beta}R_{\alpha\beta\gamma\mu}\upsilon^{\gamma}-\frac{1}{2}Q_{\mu\alpha\beta}t^{\alpha\beta} \label{eq1}
\eeq
\beq
\frac{D H_{\mu}{}^{\nu}}{d \lambda}=2 (P_{\mu}\upsilon^{\nu}-t^{\nu}{}_{\mu}) \label{Heq} 
\eeq
or
\beq
\frac{D H^{\mu\nu}}{d \lambda}=2(P^{\mu}\upsilon^{\nu}-t^{\mu\nu})+H^{\beta\nu}Q_{\alpha\beta}{}{}^{\mu}\upsilon^{\alpha} \label{Hconlaw}
\eeq
where $P_{\mu}$ is the canonical momentum,
\beq
P_{\mu}:=\frac{\partial L}{\partial \upsilon^{\mu}}
\eeq
and $H_{\mu}{}^{\nu}$  and $t^{\mu\nu}$ are the hypermomentum and the energy-momentum excitations, defined in the Lagrangian formulation through

\beq
H_{\nu}{}^{\mu}:=2 e_{\nu}{}^{a} \frac{\partial L}{\partial \dot{e}_{\mu}{}^{a}}\;\;, \;\; t^{\mu\nu}:=2 \frac{\partial L}{\partial g_{\mu\nu}}
\eeq
respectively. The hypermomentum current splits into its 3 irreducible pieces of spin (antisymmetric) dilation (trace) and shear (symmetric traceless), according to \cite{hehl1995metric}
\beq
H^{\mu\nu}=S^{\mu\nu}+\frac{H}{n}g^{\mu\nu}+\Sigma^{\mu\nu}
\eeq
where
\begin{align}    
S^{\mu\nu}&:=H^{[\mu\nu]} \;\; \;(spin) \\
H&:=H^{\mu\nu}g_{\mu\nu} \;\;\; (dilation)
\\
\Sigma^{\mu\nu}&:=H^{(\mu\nu)}-\frac{H}{n}g^{\mu\nu} \;\;\;(shear)
\end{align}
 As usual, we have the notion of kinematical (or monopole) rest mass of the particle
\beq
m:=-P_{\mu}\upsilon^{\mu} \label{kinmass}
\eeq
and the dynamical rest mass
\beq
\mathcal{M}^{2}:=-P_{\mu}P^{\mu} \label{dynmass}
\eeq
The latter being the physical one.
Let us recall that before imposing a spin supplementary condition, neither of them is a constant of motion for the standard form of the Mathisson-Papapetrou equations; a spin supplementary condition is also needed to fix the centroid of the body and close the system of equations (for a detailed review see \cite{costa2015center}). It should be noted also that along with the spin supplementary condition, the  curve parameter $\lambda$ must also identified in order to fully close the system of equations. If we identify $\lambda$ with proper time $\tau$ then $\upsilon^{\mu}\equiv u^{\mu}=d x^{\mu}/d \tau$ identifies with the particle's 4-velocity\footnote{Or n-velocity in  general.}. In order to see how the above masses evolve and under what circumstances are constants of motion, in the general case where both torsion and nonmetricity are present, it is worth recalling some well known facts about these masses in the simple case of Mathisson-Papapetrou system in Riemannian geometry.

\subsection{Constants of motion in Mathisson-Papapetrou Theory}

In Riemannian geometry, the Mathisson-Papapetrou equations read
\beq
\dot{P}^{\mu}=\frac{1}{2}S^{\alpha\beta}R_{\alpha\beta\gamma\mu}\upsilon^{\gamma} \label{MP1}
\eeq
\beq
\dot{S}^{\mu\nu}=P^{\mu}\upsilon^{\nu}-P^{\nu}\upsilon^{\mu} \label{MP2}
\eeq
Let us consider an arbitrary parametrization for the moment, so that 
\beq
\upsilon^{2}=\upsilon_{\mu}\upsilon^{\mu}=-l^{2}(x) \label{l2}
\eeq
namely that the velocity field is not necessarily normalized to (-) unity. Consequently, by differentiation we get
\beq
 \upsilon_{\mu}\dot{\upsilon}^{\mu} =-l\dot{l}
\eeq
where the metricity condition is used in this case, since the background is Riemannian.
To show how the spin supplementary conditions provide the mass constancy, for each case, we compute the derivative of each of the masses. Before doing so, let us list some useful relations that follow directly from contractions of the above:
\begin{align}
\upsilon_{\mu}\dot{S}^{\mu\nu} & = -m \upsilon^{\nu}+l^{2}P^{\mu} \label{P}\\
  P_{\mu}\dot{S}^{\mu\nu} &= -\mathcal{M}^{2}\upsilon^{\nu}+m P^{\nu} \label{PSmn} \\
  \dot{P}^{\mu} \upsilon_{\mu}&=0 \label{Pu}
\end{align}

We shall start with the kinematical rest mass. Differentiating (\ref{kinmass}), and employing (\ref{Pu}) we get
\beq
-\dot{m}=P_{\mu}\dot{\upsilon}^{\mu}
\eeq
Using (\ref{P}) it follows that
\beq
P_{\mu}\dot{\upsilon}^{\mu}=-m\frac{\dot{l}}{l}+\frac{1}{l^{2}}\upsilon_{\nu}\dot{\upsilon}_{\mu}\dot{S}^{\mu\nu}
\eeq
and now write
\beq
\upsilon_{\nu}\dot{\upsilon}_{\mu}\dot{S}^{\mu\nu}=\dot{\upsilon}_{\mu}\frac{D}{d \lambda}\Big( \upsilon_{\nu}S^{\nu\mu} \Big)-\dot{\upsilon}_{\mu}\dot{\upsilon}_{\nu}S^{\nu\mu}=\dot{\upsilon}_{\mu}\frac{D}{d \lambda}\Big( \upsilon_{\nu}S^{\nu\mu} \Big)
\eeq
since $S^{\mu\nu}$ is antisymmetric. Consequently, the kinematical rest mass evolves as
\beq
\dot{m}=m \frac{\dot{l}}{l}+\frac{\dot{\upsilon}_{\mu}}{l^{2}}\frac{D}{d \lambda}\Big( \upsilon_{\nu}S^{\nu\mu} \Big)
\eeq
From this it is clear that this mass is conserved if one imposes the Pirani constraint $\upsilon_{\nu}S^{\nu\mu}=0$ and use the proper time parametrization ($\lambda \equiv \tau$) so that $l=1$. For arbitrary parametrizations, only the ratio $m/l$ remains constant during the motion. Alternatively, what remains constant is the rest mass measured relative to the normalized velocity field, i.e. $\hat{m}=-P_{\mu}\hat{\upsilon}^{\mu}=P_{\mu}\upsilon^{\mu}/l =const.$ What this tells us basically, is that for arbitrary parametrizations, what is meaningful, is the 'normalized' mass or the rest mass relative to the normalized velocity. Therefore, for arbitrary parametrizations, the proper definition of the kinematical rest mass is
\beq
\hat{m}:=-\frac{P_{\mu}\upsilon^{\mu}}{l}=-P_{\mu}u^{\mu}
\eeq
Let us now focus on the evolution of the dynamical mass which is the physically relevant. Direct differentiation of (\ref{dynmass}) gives
\beq
-\mathcal{M}\dot{\mathcal{M}}=P_{\mu}\dot{P}^{\mu}
\eeq
To compute the right-hand side of the above, we start by (\ref{PSmn}) and perform a Leibniz rule to its left-hand side, to get
\beq
\frac{D}{d \lambda}\Big(P_{\mu}S^{\mu\nu} \Big)-\dot{P}_{\mu}S^{\mu\nu}=-\mathcal{M}^{2}\upsilon^{\nu}+m P^{\nu}
\eeq
Contracting this with $\dot{P}_{\nu}$ and using the antisymmetry of $S^{\mu\nu}$ and also equation (\ref{Pu})
we get
\beq
\dot{P}_{\nu}\frac{D}{d \lambda}\Big(P_{\mu}S^{\mu\nu} \Big)=m P_{\mu}\dot{P}^{\mu}
\eeq
Therefore, it follows that the dynamical mass evolves according to
\beq
\dot{\mathcal{M}}=-\frac{1}{\mathcal{M}m}\dot{P}_{\nu}\frac{D}{d \lambda}\Big(P_{\mu}S^{\mu\nu} \Big)
\eeq
from which we infer that this mass is conserved ($\mathcal{M}=const.$) if one imposes the Tulczyjew constraint
\beq
P_{\mu}S^{\mu\nu}=0
\eeq
Now, on to the squared spin magnitude, contracting (\ref{MP2}) with $S_{\mu\nu}$ it follows that
\beq
\frac{D}{d \lambda}\Big( S_{\mu\nu} S^{\mu\nu} \Big)=4S^{\mu\nu}P_{\mu}\upsilon_{\nu}
\eeq
and we see that the squared spin magnitude is a constant of motion for either the Pirani or  Tulczyjew constraints.

It is worth mentioning that for relativistic tops, this constancy of the spin magnitude follows without ever demanding  a spin supplementary condition \cite{Hanson:1974qy,hojman1976spinning}.\footnote{The same is not true however, for the kinematical and dynamical masses, whose constancy again is only guaranteed by imposing the appropriate spin supplementary condition.}  What distinguishes then the relativistic top from an arbitrary relativistic spinning body, is that for the top the constancy of the squared spin magnitude is a built-in property that does not have to be imposed separately but comes as a consequence of the equations of motion for the top. All of the above are standard well known facts. Let us now move on to the new stuff.

\section{Novel Regge-like Trajectories and Generalized Spin and Shear Supplementary Conditions}
Let us now come back to the general case of a particle possessing the full hypermomentum charge (endowed with spin and hadronic properties, e.g. color charge), and rewrite the equations as
\beq
\dot{P}_{\mu}=-2 S_{\mu\alpha\beta}\upsilon^{\alpha}P^{\beta}+\frac{1}{2}H^{\alpha\beta}R_{\alpha\beta\gamma\mu}\upsilon^{\gamma}-\frac{1}{2}Q_{\mu\alpha\beta}t^{\alpha\beta} \label{Pevo}
\eeq
\beq
\dot{H}^{\mu\nu}=2 (P^{\mu}\upsilon^{\nu}-t^{\mu\nu})+H^{\beta\nu}Q_{\alpha\beta}{}{}^{\mu}\upsilon^{\alpha} \label{Hevo2}
\eeq
Taking the antisymmetric part of the latter, there follows the generalized evolution law for spin
\beq
\dot{S}^{\mu\nu}=P^{\mu}\upsilon^{\nu}-P^{\nu}\upsilon^{\mu}+\xi^{\mu\nu} \label{Spinevo}
\eeq
where we have defined
\beq
\xi^{\mu\nu}=-\xi^{\nu\mu}:=H^{\beta[\nu}Q_{\alpha\beta}{}{}^{\mu]}\upsilon^{\alpha}
\eeq
Substituting the evolution equation (\ref{Spinevo}) back in (\ref{Hevo2}) and solving for the energy-momentum tensor, it follows that
\beq
t^{\mu\nu}=P^{\mu}\upsilon^{\nu}-\frac{1}{2}\dot{H}^{\mu\nu}+ \frac{1}{2}H^{\beta\nu}Q_{\alpha\beta}{}{}^{\mu}\upsilon^{\alpha}
\eeq
Notice that due to (\ref{Spinevo}) the right hand side of the above is properly symmetric as expected. From contractions of (\ref{Spinevo}) we get the useful relations
\begin{align}
    P_{\mu}\dot{S}^{\mu\nu}&=-\mathcal{M}^{2} \upsilon^{\nu}+m P^{\nu}+\xi^{\mu\nu}P_{\mu} \label{PS}\\
\upsilon_{\mu}\dot{S}^{\mu\nu}&=-m \upsilon^{\nu}+l^{2} P^{\nu}+\xi^{\mu\nu}\upsilon_{\mu} \label{Pu}\\
P_{\mu}\upsilon_{\nu}\dot{S}^{\mu\nu}&=l^{2}\mathcal{M}^{2}-m^{2}+\xi^{\mu\nu}P_{\mu}\upsilon_{\nu}
\end{align}
Let us also note that due to non-metricity the derivatives of the magnitudes of vectors receive an additional contribution. For the cases of the magnitudes of $P^{\mu}$ and $\upsilon^{\mu}$, by differentiation we find
\begin{align}
\mathcal{M}\dot{\mathcal{M}}&=-\dot{P}_{\mu}P^{\mu}-\frac{1}{2}Q_{\alpha\mu\nu}\upsilon^{\alpha}P^{\mu}P^{\nu}  \label{MM}\\
l\dot{l}&=-\dot{\upsilon}_{\mu}\upsilon^{\mu}-\frac{1}{2}Q_{\alpha\mu\nu}\upsilon^{\alpha}\upsilon^{\mu}\upsilon^{\nu} \label{ll}
\end{align}
Additionally, contraction of (\ref{Pevo}) with $\upsilon^{\mu}$ now gives
\beq
\dot{P}_{\mu}\upsilon^{\mu}=-\frac{1}{2}Q_{\alpha\mu\nu}\upsilon^{\alpha}t^{\mu\nu}
\eeq
We are now in a position to find the evolution laws for the masses. We start with the dynamical mass. To this end, we contract (\ref{PS}) with $\dot{P}_{\nu}$  and use equations (\ref{MM}) , to arrive at
\beq
\boxed{\dot{P}_{\nu}\left[ \frac{D}{d \lambda}\Big( P_{\mu}S^{\mu\nu}\Big)-\xi^{\mu\nu}P_{\mu} \right]=\frac{1}{2}Q_{\alpha\mu\nu}\upsilon^{\alpha}\Big( \mathcal{M}^{2}t^{\mu\nu}-mP^{\mu}P^{\nu}\Big)-m \mathcal{M}\dot{\mathcal{M}}} \label{dynmassevo}
\eeq
This is the (dynamical) mass evolution equation in the most general case.
From the above it is obvious that even with the imposition of the Tulczyjew constraint $S^{\mu\nu}P_{\mu}=0$, along with the constraint
\beq
\xi^{\mu\nu}P_{\mu} =0
\eeq
the dynamical mass cannot  be a constant of motion in general, due to the nonmetricity of space. Note, however, that in the pole approximation $t^{\mu\nu}\approx m U^{\mu}U^{\nu}$, with $P^{\mu}=\mathcal{M}U^{\mu}$, the dynamical mass is a constant of motion.

By eliminating $t^{\mu\nu}$ from the above expression we can also  derive the alternative form
\begin{gather}
    m \mc \dot{\mc}+\frac{1}{4}\mc^{2}\y^{\alpha}Q_{\alpha\nu}{}{}^{\mu}\dot{H}_{\mu}{}^{\nu}=-\dot{P}_{\nu}\left[ \frac{D}{d \lambda}\Big( P_{\mu}S^{\mu\nu}\Big)-\xi^{\mu\nu}P_{\mu} \right]\nonumber \\+\frac{1}{2}\y^{\alpha}Q_{\alpha\nu}{}{}^{\mu}P_{\mu}(\dot{S}^{\nu\lambda}P_{\lambda}-\xi^{\nu\lambda}P_{\lambda})
\end{gather}

In a similar manner, starting from (\ref{Pu}) we find the evolution of the kinematical mass to be
\begin{gather}
\dot{\upsilon}_{\nu}\left[ \frac{D}{d \lambda}\Big( \upsilon_{\mu}S^{\mu\nu}\Big)-\xi^{\mu\nu}\upsilon_{\mu} \right]=m l \dot{l}-l^{2}\dot{\left( \frac{m}{l} \right)} \nonumber \\
+\frac{1}{2}Q_{\alpha\mu\nu}\upsilon^{\alpha}\Big(m \upsilon^{\mu}\upsilon^{\nu}-2 l^{2}P^{\mu}P^{\nu}+l^{2}t^{\mu\nu}\Big)  \label{kinmassevo}
\end{gather}
which, as expected, is also non-constant even for a generalized version of the Pirani constraint $S^{\mu\nu}\y_{\mu}=0$.

\subsection{Regge-like trajectories with pure dilation charge}

Let us now constrain the nature of the mictrostructured body and study the case where the latter is charged only under dilation. Physically, this is seen as an elastic breathing mode which quantum-mechanically  corresponds to changes in the excitation levels of atoms or molecules.  In this case, the hypermomentum is a pure trace, $H_{\mu}{}^{\nu}=H \delta_{\mu}^{\nu}$ and from (\ref{Heq}) we  extract the metrical energy-momentum tensor
\beq
t^{\mu\nu}=P^{\mu}\upsilon^{\nu}-\frac{\dot{H}}{2}g^{\mu\nu} \label{tmndil}
\eeq
Then, given that $t^{\mu\nu}$ is symmetric, taking the antisymmetric part of the above gives
\beq
0=P^{[\mu}\upsilon^{\nu]}
\eeq
meaning that $P^{\mu}$ is parallel to $\upsilon^{\mu}$. Given the magnitudes (\ref{l2}) and (\ref{dynmass}), the proportonality factor is fixed to $\mathcal{M}/l$ and so the momentum-velocity relation is the usual one
\beq
P^{\mu}=\mathcal{M}\frac{\upsilon^{\mu}}{l}=\mathcal{M} u^{\mu} \label{Pdil}
\eeq
Note also that in this case, the kinematical and dynamical masses are proportional $m=\mathcal{M}l$ and identify for proper time parametrization, i.e. $m \equiv \mathcal{M}$ for $\lambda \equiv \tau$. By substituting the above result (\ref{Pdil}) back to equation (\ref{tmndil}), the latter becomes
\beq
t^{\mu\nu}=\frac{\mathcal{M}}{l}\upsilon^{\mu}\upsilon^{\nu}-\frac{\dot{H}}{2}g^{\mu\nu} =\frac{l}{\mathcal{M}}P^{\mu}P^{\nu}-\frac{\dot{H}}{2}g^{\mu\nu}  \label{tmndil2}
\eeq
Let us now investigate the evolution of the rest mass.  Setting spin to zero in (\ref{dynmassevo}) and using the relation $m=\mathcal{M}l$ we get
\beq
\mathcal{M}\dot{\mathcal{M}}=\frac{1}{2}\upsilon^{\alpha}Q_{\alpha\mu\nu}\left(\frac{\mathcal{M}}{l}t^{\mu\nu}-P^{\mu}P^{\nu} \right)
\eeq
and upon using (\ref{tmndil2}), we arrive at
\beq
\boxed{\dot{M}=-\frac{Q_{\mu}\upsilon^{\mu}}{4 l}\dot{H}} \label{ReggeH}
\eeq
Upon formal integration this leads to a novel \underline{Regge-trajectory} involving the mass and the dilation charge
\beq
\mathcal{M}(H)=-\int Q_{\mu}u^{\mu}H' d\tau+\mathcal{M}_{0} \;\;, \;\; H'=\frac{d H}{d \tau}
\eeq
A physical interpretation is  now in order.  A dilation charged mictrostructured particle interacts with the background non-metricity causing its mass to fluctuate. On a quantum level this is interpreted as a change in the excitation level of the structure (be it a nucleus, atom or even molecule) due   
to the microstretching of the host space. In this microscopic interpretation, this corresponds to changing the principal quantum number (n) while keeping the orbital angular momentum (l) fixed, leading to radial excitations (breathing modes). 
Going one step further one could associate this mass shift with the so-called Roper resonance \cite{Roper1964}.\footnote{The Roper resonance appears to be a heavier exact copy of the proton, with a mass $50\%$ heavier. For over 50 years the explanation of its mass shift  remained a puzzle that could not be addressed by QCD \cite{Burkert:2017djo}. After a long effort, it is now established that the Roper resonance is proton's first radial excitation \cite{Burkert:2017djo,Burkert:2025aze}. Here we argue that this excitation can be explained by the interaction with an enriched spacetime structure, i.e. the micro-stretching of spacetime.} In particular one could argue that the Roper resonance appears due to the interaction of the nuclear structure with the non-metricity. This conclusion is supported by the fact that non-metricity can be probed by the hadronic properties of matter \cite{hehl1997ahadronic}.

Note that  in the pole approximation where the dilation charge is constant \cite{Iosifidis:2023eom}, the rest mass of the particle does become a constant of motion. Since the coupling involves only the Weyl vector another case where the mass remains constant, without any approximation, is when the nonmetricity is trace free (i.e. the Weyl vector $Q_{\mu}$ vanishes).

Performing a post-Riemannian expansion to the momentum evolution equation and using the above results, the latter takes the form\footnote{Here $\frac{\widetilde{D}}{d \lambda}:=\y^{\mu}\widetilde{\nabla}_{\mu}$ where $\widetilde{\nabla}_{\mu}$ is the Levi-Civita covariant derivative. }
\beq
\frac{\widetilde{D} P^{\alpha}}{d \lambda}=\frac{1}{4}Q^{\alpha}\dot{H}+\frac{1}{2}H\upsilon_{\beta}\partial^{[\beta}Q^{\alpha]} \label{DtildeP}
\eeq
Interestingly, this equation is identical to that of a charged particle in an Electromagnetic field with variable charge, where the role of electric charged is played by the dilation charge here and the external field is that of nonmetricity. It is also worth remarking that the relation (\ref{ReggeH}) is mathematically equivalent to the mass-charge relation one typically gets in brane-world and Kaluza-Klein scenarios, see e.g. \cite{PoncedeLeon:2002xr}. Finally, further substitution of (\ref{Pdil}) and (\ref{ReggeH}) into (\ref{DtildeP}), gives 
\beq
\mathcal{M}\frac{\widetilde{D}}{d \lambda}\left( \frac{\upsilon^{\alpha}}{l}\right)=\frac{1}{4}\left( Q^{\alpha}+\frac{1}{l^{2}}(Q_{\mu}\upsilon^{\mu})\upsilon^{\alpha}\right)\dot{H}+\frac{1}{2}H\upsilon_{\beta}\partial^{[\beta}Q^{\alpha]}
\eeq
exhibiting a non-geodesic motion due to dilation as we explained above.
 \subsubsection{Scale Invariant Case}

The precise form of the Regge-like trajectory and consequently the mass depends on the specific properties of matter under consideration. A physically relevant example where the mass is computed exactly is that of a scale invariant matter. In this case $t=0$ and from (\ref{tmndil}) we infer that
\beq
\dot{H}=-\frac{2 l \mathcal{M}}{n} \label{dildot}
\eeq
which we may substitute in (\ref{ReggeH}) to find
\beq
\dot{\mathcal{M}}+\frac{Q_{\mu}\upsilon^{\mu}}{2 n}\mathcal{M}=0
\eeq
 which, upon trivially integrated, provides the explicit mass relation
 \beq
\mc =\mc_{0}e^{-\frac{1}{2n}\int Q_{\mu}\upsilon^{\mu}d \lambda} .\label{massform}
 \eeq
We see that when the particle enters a non-metric region of spacetime its initial mass $\mc_{0}$ starts to shift. Substituting the latter into (\ref{dildot}) and integrating gives us the form of the dilation charge:
\beq
H=C_{0}-\frac{2 \mc_{0}}{n}\int e^{-\frac{1}{2 n}\int Q_{\mu}d x^{\mu}} d \tau \;\;, \;\; C_{0}=const.
\eeq
Upon specification of the background geometry (i.e. knowledge of $Q_{\mu}$ in this case) the explicit Regge trajectory relation $\mc (H)$ can be obtained. To give an example for $Q_{0}=1$ and $Q_{i}=0$ for $i=1,2,3$ we find the explicit relation
\beq
\mc(H)=\mc_{0}+\frac{1} {4}(H-H_{0})
\eeq
 where $\mc_{0}$ and $H_{0}$ are respectively the initial mass and dilation charge.

\subsection{Regge-like Trajectories with Color Charge}
Let us now ignore the spin and dilation charge of the particle and focus only in its color charge as comes from the pure shear sector of hypermomentum (i.e set $H^{\mu\nu}\equiv \Sigma^{\mu\nu}$). 
The very vanishing of spin, gives us in this case
\beq
P^{[\mu}\y^{\nu]}=-\xi^{\mu\nu}=\Sigma^{\beta[\mu}\mu_{\beta}{}^{\nu]} \label{Pshear}
\eeq
where we have defined\footnote{As a matter of fact, in exterior form language these $\mu_{ab}$ are nothing else than the nonmetricity components. Indeed, in exterior calculus nonmetricity 1-form is defined by $Q_{ab}=-D g_{ab}$, $a,b$ representing anholonomic indices, and therefore $Q_{ab}=Q_{\mu ab}dx^{\mu}=\mu_{ab}d \lambda$, confirming that $\mu_{ab}$ are the components of nonmetricity along the particle's trajectory.}
\beq
\mu_{\alpha\beta}=\mu_{\beta\alpha}:=\y^{\gamma}Q_{\gamma\alpha\beta}
\eeq
Contracting (\ref{Pshear}) with $\y^{\nu}$ we find the momentum-velocity relation
\beq
P^{\mu}=\frac{m}{l^{2}}\y^{\mu}+\frac{1}{l^{2}}\xi^{\mu\nu}\y_{\nu}
\eeq
where  $\xi^{\mu\nu}=-\Sigma^{\beta[\mu}\mu_{\beta}{}^{\nu]}$ in this case. Physically, the above expression confirms the known result that anisotropic shear deformations (for instance in a nucleus) contribute to the total momentum. Of course this is not the case for (isotropic) radial excitations and in that case momentum and velocity are parallel (see previous section on pure dilation).  Let us note also that in the pure shear case we  have the identity
\beq
\dot{\Sigma}_{\lambda}{}^{[\mu}g^{\nu]\lambda}=\Sigma^{\beta[\mu}\mu_{\beta}{}^{\nu]}
\eeq
Using the above relations the mass evolution equation (\ref{dynmassevo}) becomes
\begin{gather}
    m\mc \dot{\mc}-\dot{P}_{\nu}\xi^{\mu\nu}P_{\mu}=\nonumber \\
  \hat{\mu}_{\mu\nu}  \frac{1}{2}\mc^{2}\left[ \frac{1}{2}(\Sigma^{\beta\nu}\mu_{\beta}{}^{\mu}-\dot{\Sigma}^{\mu\nu})+\frac{1}{m}\xi^{\lambda\nu}\y_{\lambda}P^{\mu} \right] \label{masseqshear}
\end{gather} 
where 
\beq
\hat{\mu}_{\mu\nu}=\mu_{\mu\nu}-\frac{\mu}{n}g_{\mu\nu}=\y^{\alpha}\Big( Q_{\alpha\mu\nu}-\frac{Q_{\alpha}}{n}g_{\mu\nu}\Big)
\eeq
is properly traceless ($\hat{\mu}=g^{\alpha\beta}\hat{\mu}_{\alpha\beta}=0$), 
namely only the traceless part of non-metricity contributes, as a result of the vanishing dilation.

Similarly to the spin case, we now need to impose a 'Shear Supplementary Condition'. Perhaps the most obvious choice, having the spin case as a guide, is to impose either\footnote{In general, one might feel tempted to impose a constraint on the full hypermomentum like $H^{\mu\nu}P_{\mu}=0$ or $H^{\mu\nu}u_{\mu}=0$ as a generalizations of the Tulczyjew and Pirani constraints as was done in \cite{obukhov1993hyperfluid}. However, one easily realizes  that such  constraints immediately lead to inconsistencies, since for the pure dilation case these translate to vanishing momentum for the former and vanishing velocity for the latter. Here, we shall avoid the imposition of such unphysical constraints.}
\beq
\xi^{\mu\nu}P_{\mu}=0
\eeq
or 
\beq
\xi^{\mu\nu}\y_{\mu}=0
\eeq
the former being the analogue of the  Tulczyjew constraint and the latter resembling the Pirani constraint. It is straightforward to show that these two conditions are logically equivalent; imposing either one necessitates the other. Furthermore, they also imply that
\beq
P^{\mu}=\frac{\mathcal{M}^{2}}{m}\y^{\mu}
\eeq
and $m=\mc l$, telling us that the masses identify for proper time parametrizations. With these results, eq. (\ref{masseqshear}) simplifies greatly and takes the nice form
\beq
\boxed{\dot{\mc}=-\frac{1}{4l}\hat{\mu}^{\mu}{}_{\nu}\dot{\Sigma}_{\mu}{}^{\nu}} \label{ReggeS}
\eeq
Quite remarkably, this is structurally identical to (\ref{ReggeH}) by replacing the dilation part with that of shear and also the trace part of nonmetricity with its trace-free counterpart. As a matter of fact one equation maps to the other by simultaneously exchanging  $\dot{H}\delta_{\mu}^{\nu} \leftrightarrow \dot{\Sigma}_{\mu}{}^{\nu}$ and $Q_{\alpha}\y^{\alpha}\delta_{\nu}{}^{\mu}\leftrightarrow \hat{\mu}^{\mu}{}_{\nu}$. 
Therefore there is a duality between dilation-Weyl,  and Shear-Traceless nonmetricity couplings. Our findings  clearly show that for minimally coupled matter\footnote{The picture drastically changes if one considers non-minimally coupled matter. Then the conservation laws change \cite{Puetzfeld:2014qba} and it is possible to have spin current generate nonmetricity or dilation and shear charges generating torsion, see Appendix of \cite{Iosifidis:2020gth} for such an example and also \cite{Puetzfeld:2013sca}.} the dilation only couples to the trace part of nonmetricity while the shear couples to its traceless counterpart. This is in perfect 
agreement with previous claims \cite{hehl1995metric}.

\subsection{Generalization to particles carrying both dilation and shear charges and the  'Shear Supplementary Condition'}

Let us now generalize the reults of the previous two subsections and consider a particle carrying both dilation and shear charges and no spin. For vanishing spin, the dynamical mass evolution equation (\ref{dynmassevo}) reduces to
\begin{gather}
    m \mc \dot{\mc}+\frac{1}{4}\mc^{2}\y^{\alpha}Q_{\alpha\nu}{}{}^{\mu}\dot{H}_{\mu}{}^{\nu}=-\frac{1}{2}P_{\mu}\Big( 2\dot{P}_{\nu}\xi^{\mu\lambda}+\mu^{\mu}{}_{\nu}\xi^{\nu\lambda}P_{\lambda} \Big)
\end{gather}
where now $\xi^{\mu\nu}=\Sigma^{\beta[\nu}\mu_{\beta}{}^{\mu]}$ due to the vanishing of spin. We see therefore that imposing the \underline{'Shear Supplemetary Condition'}:
\beq
\xi^{\mu\nu}P_{\mu}=0 \label{ShearSC}
\eeq
gives Regge-like trajectories carrying both dilation and shear charges:
\beq
\dot{\mc}+\frac{1}{4l }\mu^{\mu}{}_{\nu}\dot{H}_{\mu}{}^{\nu}=0 \label{ReggeSD}
\eeq
where $H_{\mu}{}^{\nu}=\Sigma_{\mu}{}^{\nu}+\frac{1}{n}\delta_{\mu}^{\nu}H$.
In extracting (\ref{ReggeSD}) we have used the fact that 
for vanishing spin and the imposition of the above shear supplementary condition, it follows from (\ref{PS}) that
\beq
P^{\mu}=\frac{\mc^{2}}{m}\y^{\mu}
\eeq
We conclude, therefore, that for vanishing spin the shear supplementary condition (\ref{ShearSC}) also guarantees that momentum remains parallel to the velocity.

\subsubsection{Conserved Dilation and Shear  Sector }

In certain cases, the systematics of hadrons indicate an approximate conservation of
intrinsic hypermomentum \cite{NeemanSijacki1979}. The approximate conservations for the dilation and shear charges read 
\beq
\frac{D H}{d \lambda}\approx0\;\;, \;\;\frac{D \Sigma_{\mu}{}^{\nu}}{d \lambda}\approx 0
\eeq
Then from (\ref{ReggeSD}) it immediately follows that in such instances the mass is approximately conserved:
\beq
\dot{\mc} \approx 0
\eeq

\subsection{Riemann-Cartan-Weyl Geometry}
In the previous subsection, we constrained the microstructure of the particle, leaving the background geometry totally generic. We shall now leave all the charges (spin, dilation, and shear) of the particle active and constrain the underlying geometry. A geometrically interesting and popular subspace of the generic Metric-Affine Geometry is the so-called Riemann-Cartan-Weyl space where curvature and torsion are general but nonmetricity is taken to be of a pure trace (Weyl) type, $Q_{\alpha\mu\nu}=\frac{Q_{\alpha}}{n}g_{\mu\nu}$. In this case the mass evolution equation (\ref{dynmassevo}) becomes
\beq
\dot{P}_{\nu}\left[ \frac{D}{d \lambda}\Big( P_{\mu}S^{\mu\nu}\Big)-S^{\mu\nu}P_{\mu} \right]=\frac{1}{2n}Q_{\alpha}\upsilon^{\alpha}g_{\mu\nu}\Big( \mathcal{M}^{2}t^{\mu\nu}-mP^{\mu}P^{\nu}\Big)-m \mathcal{M}\dot{\mathcal{M}} \label{MassRCW}
\eeq
from which we see that the imposition of the usual Tulczyjew constraint $P_{\mu}S^{\mu\nu}=0$ brings it to the form
\beq
\frac{\mathcal{M}}{2 n}Q_{\alpha}\upsilon^{\alpha}(t+m)=m\dot{\mathcal{M}}
\eeq
But, contracting (\ref{tmndil}) in $\mu=\nu$ it follows that $t+m=-1/2 \dot{H}$ and consequently the latter reads
\beq
\dot{\mathcal{M}}+\frac{\mathcal{M}}{4n m}Q_{\alpha}\upsilon^{\alpha}\dot{H}=0
\eeq
Quite remarkably, we see a duality with the pure dilation result (in the generic metric-affine space) of the previous subsection, given that one exchanges
\beq
l \mathcal{M} \leftrightarrow nm
\eeq
We also conclude that if the body under consideration does not possess a dilation charge, its rest mass will be a constant of motion, even if the body is charged under spin and shear.

\section{Conclusions}

We studied the dynamical mass evolution for spinning hypermomentum charged (micro-structured) particles moving in generalized non-Riemannian backgrounds. In full generality, the evolution of the mass is given by eq. (\ref{dynmassevo}) which shows that in generic situations the mass is not a constant of motion even with the imposition of a spin supplementary condition. Furthermore, focusing on the pure dilation and pure shear cases, we derived generalized Regge-like trajectories that relate the particle's rest mass to the dilation and shear charges. These new Regge trajectories are given by equations (\ref{ReggeH}) and (\ref{ReggeS}) respectively, and are further generalized to (\ref{ReggeSD}).
A specific example where an explicit form of the mass relation can be extracted, was also given. 

Finally we introduced the notion of a Shear Supplementary Condition, in direct analogy to the Spin Supplementary Condition, and showed how this condition can, in certain situations, render mass to be a constant of motion even in the presence of nonmetricity.

\section{Acknowledgments}

  This work was supported by the  Istituto Nazionale di Fisica Nucleare (INFN), Sezioni  di Napoli,  {\it Iniziative Specifiche} QGSKY. I would like to thank very much Friedrich Hehl  for many useful comments and discussions.

\bibliographystyle{unsrt}
\bibliography{ref}

\end{document}